\documentstyle[12pt,cite,epsfig]{article}
\let\LARGE=\Large
\let\Large=\large
\let\large=\normalsize
\newcommand{\Abstract}[1]{\begin{center}{\large ABSTRACT}\end{center}{#1}
\vskip 0.4in}

\def\Thebibliography#1{\section*{References}\list
 {[\arabic{enumi}]}{\settowidth\labelwidth{[#1]}\leftmargin\labelwidth
 \advance\leftmargin\labelsep
 \usecounter{enumi}}
 \def\newblock{\hskip .11em plus .33em minus .07em}
 \sloppy\clubpenalty4000\widowpenalty4000
 \sfcode`\.=1000\relax}

\def\Journal#1&#2&#3(#4){\unskip, #1~{#2} (#4) #3}

\def\NPB{Nucl.\ Phys.\ B}

\def\PRL{Phys.\ Rev.\ Lett.}

\def\BBB{\stackrel{\scriptscriptstyle
    {(-)\hphantom{\scriptstyle o}}}{\rm B^o}}
\def\ffb{\stackrel{\scriptscriptstyle (-)}{f}}

\newcommand{\be}{\begin{equation}}
\newcommand{\ee}{\end{equation}}
\newcommand{\ba}{\begin{array}{c}}
\newcommand{\ea}{\end{array}}
\newcommand{\beqn}{\begin{eqnarray}}
\newcommand{\eeqn}{\end{eqnarray}}

\newcommand{\bi}{\begin{itemize}}
\newcommand{\ei}{\end{itemize}}

\newcommand{\cL}{{\cal L}}

\newcommand{\cO}{{\cal O}}
\newcommand{\cP}{{\cal P}}

\newcommand{\cJ}{{\cal J}}

\newcommand{\rms}{\rm\scriptsize}
\begin{document}

%
%

\begin{titlepage}
\vspace{0.3in}
\begin{flushright}
{CERN-TH.7114/93}
\end{flushright}
\vspace*{1.6cm}
\begin{center}
{\LARGE \bf CP VIOLATION}

\vspace*{0.4cm}
{\bf A. Pich}$^{*\dagger}$

Theory Division, CERN, CH-1211 Geneva 23

\end{center}
\vspace*{1.5cm}


\Abstract{An overview of the phenomenology of CP violation is presented.
The Standard Model mechanism of CP violation and its main experimental
tests, both in the kaon and bottom systems, are discussed.}

\vspace*{2.2cm}
\begin{center}
{Lectures given at the\\
ICTP Summer School in High Energy Physics and Cosmology \\
ICTP, Trieste, Italy, June 1993 \\
and at the \\
Escuela Latinoamericana de F\'{\i}sica (ELAF'93)\\
Mar del Plata, Argentina, July 1993}
\end{center}

\vspace*{2.5cm}

{\footnotesize\baselineskip 10pt \noindent
$^*$ On leave of absence from
Departament de F\'{\i}sica Te\`orica, Universitat de Val\`encia,
and IFIC, Centre Mixte Universitat de Val\`encia--CSIC,
E-46100 Burjassot, Val\`encia, Spain.}

{\footnotesize\baselineskip 10pt \noindent
$^\dagger$ Work supported in part
by CICYT (Spain), under grant No. AEN93-0234.}

\vfill
\begin{flushleft}
{CERN-TH.7114/93 \\
December 1993}
\end{flushleft}
\end{titlepage}

%


\title{\bf CP VIOLATION}
\author{A. PICH\thanks{On leave of absence from
Departament de F\'{\i}sica Te\`orica, Universitat de Val\`encia,
and IFIC, Centre Mixte Universitat de Val\`encia--CSIC,
E-46100 Burjassot, Val\`encia, Spain.}\ \thanks{Work supported in part
by CICYT (Spain), under grant No. AEN93-0234.}
\\ Theory Division, CERN, CH-1211 Geneva 23}
\date{}
\maketitle 
\Abstract{An overview of the phenomenology of CP violation is presented.
The Standard Model mechanism of CP violation and its main experimental
tests, both in the kaon and bottom systems, are discussed.}
\section{Introduction}
\label{sec:introduction}

Charge conjugation (C) and Parity (P) are drastically violated by the
weak interactions; however, their product CP happens to be a good
symmetry in nearly all observed phenomena.
So far, only in the decay
of neutral kaons a slight violation of the CP symmetry
($\sim 2\times 10^{-3}$)
has been established.
No observation of this phenomena has been
made in any other system.
Our understanding of CP violation
is therefore very poor. We do not know
yet whether CP violation is simply an accident proper to the neutral kaons,
due to the fact that the $K^0$ oscillates into its antiparticle,
or if it is a general property of weak interactions which could
manifest in other weak decays.

In the three-generation Standard Model (SM), CP violation
originates from the single phase  naturally occurring in the
Cabibbo-Kobayashi-Maskawa (CKM) quark-mixing matrix \cite{ko:1,ca:1}.
The present experimental observations are in agreement with the SM
expectations; nevertheless, the correctness of the CKM mechanism is
far from being proved. We have no
understanding of why nature has chosen the
number and properties of fundamental fields just so that
CP-violation may be possible.
Like fermion masses and quark-mixing angles, the origin of the CKM phase
lies in the more obscure part of the SM Lagrangian: the scalar sector.
Obviously, CP violation could well be a sensitive probe for new
physics beyond the SM.

The purpose of these lectures is to give an overview
of the phenomenology of CP violation.
The SM mechanism of CP violation is presented in Sect. \ref{sec:SM}.
Sect. \ref{sec:mechanisms} shows different ways of generating an
observable CP-violating effect and
summarizes the present experimental evidence in the kaon
system;
the SM predictions for these CP-violation observables are also discussed.
The strong CP problem is very briefly mentioned in Sec. \ref{sec:strong_CP}.
Sects. \ref{sec:rare} and \ref{sec:bottom} describe future tests
of CP violation in rare $K$ decays and with $B$ mesons, respectively.
The present tests on the unitarity of the CKM matrix and
a few sumarizing comments are finally collected in
Sect. \ref{sec:summary}.
A more extensive discussion can be found in many reviews of
CP violation, published recently
\cite{ref:CP_book,ref:CP_reviews,ref:bh92,ref:nakada,ref:CP_talks}.

\section{The CKM Mechanism of CP Violation}
\label{sec:SM}

CP violation requires the presence of complex phases. The only part of the SM
Lagrangian containing complex couplings is the Yukawa sector,
introduced to generate the femion masses:
\be\label{eq:lyukawa}
{\cal L}_{Yukawa} = -\left(
\overline{U}^\prime_L \ {\bf m}\ U^\prime_R\ +\
\overline{D}^\prime_L \ \widetilde{\bf m}\ D^\prime_R\
+\ h.c.\right)
\left( 1 + {\Phi_o \over v} \right) .
\ee
Here $\Phi_o$ stands for the scalar Higgs field
and $v$ is its vacuum expectation value.
The quark fields $U^\prime_{L,R}$ and $D^\prime_{L,R}$
are the 3-component vectors in flavour space
for the up- and down-type quarks respectively,
\be
U^\prime_{L,R} = \left( {1\mp \gamma_5 \over 2} \right)
\left( \begin{array}{c} u^\prime \\ c^\prime \\ t^\prime \end{array} \right)
\, , \qquad \qquad \qquad
D^\prime_{L,R} = \left( {1\mp \gamma_5 \over 2} \right)
\left( \begin{array}{c} d^\prime \\ s^\prime \\ b^\prime \end{array} \right)
\, ,
\ee
and ${\bf m}$, $\widetilde{\bf m}$ are $3\times 3$ mass matrices of
arbitrary complex numbers,
the elements of which are $m_{ij} = -{v \over \sqrt{2} } Y_{ij}$
and $\widetilde{m}_{ij} = -{v \over \sqrt{2} } \widetilde{Y}_{ij}$
where $Y_{ij}$ and $\widetilde{Y}_{ij}$ are the Yukawa coupling constants
with $i \equiv u,c$ or $t$ and $j \equiv d,s$ or $b$.

In general, ${\bf m}$ and $\widetilde{\bf m}$ are not diagonal.
The diagonalization of the quark-mass matrices,
${\bf m}_D = V_L{\bf m} V^{\dagger}_R$
and $\widetilde{\bf m}_D = \widetilde{V}_L \widetilde{\bf m}
\widetilde{V}^{\dagger}_R$,
defines the physical (mass-eigenstates)
fermion fields $U_{L,R} = V_{L,R} U^\prime_{L,R}$
and $D_{L,R} = \widetilde{V}_{L,R} D^\prime_{L,R}$
where $V_{L,R}$ and $\widetilde{V}_{L,R}$ are unitary matrices.

The coupling of the physical quarks to
the neutral $Z$ preserves the observed absence
of flavour-changing neutral currents (FCNC),
while their coupling to the charged $W^{\pm}$ introduces
the mixing between families.
The charged-current couplings are:
\be
{\cal L}_W = {g \over \sqrt{2}} \left\{
\overline{U}_L \gamma^{\mu} W^+_{\mu}{\bf V}D_L\ +\
\overline{D}_L \gamma^{\mu }W^-_{\mu} {\bf V}^{\dagger}U_L\
\right\}  ,
\ee
where ${\bf V}\equiv V_L \widetilde{V}^{\dagger}_L$
is an unitary $3\times 3$ matrix called
the quark-mixing or Cabibbo-Kobayashi-Maskawa (CKM) matrix
\cite{ko:1,ca:1}:
\be
{\bf V} \ \ =\ \ \
\left[ \matrix{\displaystyle V_{ud} & \displaystyle \ \ V_{us}
& \displaystyle \ \ V_{ub} \cr\displaystyle  & \displaystyle  & \displaystyle
\cr\displaystyle V_{cd} & \displaystyle \ \ V_{cs} & \displaystyle \ \ V_{cb}
\cr\displaystyle  & \displaystyle  & \displaystyle \cr\displaystyle V_{td} &
\displaystyle \ \ V_{ts} & \displaystyle \ \ V_{tb} \cr} \right] \, .
\ee
Obviously this picture can be extended to more than three families
and could be valid (with massive neutrinos) for the leptonic sector.

The Yukawa couplings $Y_{ij}$ and $\widetilde{Y}_{ij}$ are
completely arbitrary complex numbers; therefore, the resulting quark masses
and CKM matrix elements cannot be predicted in any way.
Our lack of understanding of the scalar sector translates into
a proliferation of free parameters.
A general unitary $n \times n$ matrix can be characterized by $n^2$
independent real parameters. Not all these parameters are, however,
physical observables.
The SM Lagrangian remains
invariant under the following transformation:
\be
U^i_{L,R}\longrightarrow e^{i\phi(u^i)} U^i_{L,R} , \quad
D^j_{L,R}\longrightarrow e^{i\phi(d^j)} D^j_{L,R} , \quad
V_{i,j}\longrightarrow e^{i\left(\phi(u^i)-\phi(d^j)\right)} V_{i,j} .
\ee
Thus, $(2n-1)$ phases, where $n$
is the number of fermion families,
can be reabsorbed by an appropriate
redefinition of the quark fields.
The most general CKM matrix contains then $(n-1)^2$ real parameters:
$n(n-1)/2$ mixing angles and $(n-1)(n-2)/2$ phases.
With more than two families, the elements of ${\bf V}$
 can be complex numbers which
allow the possibility for generating
CP violation through the interference of two diagrams involving different
matrix elements.

With $n=3$, the CKM matrix is described by 3 angles and 1 phase.
Different (but equivalent) representations can be found in the literature.
The Particle Data Group \cite{ref:PDG92}
advocates the use of the following one as the
``standard'' CKM parametrization:
\be
{\bf V}\, = \, \left[
\begin{array}{ccc}
c_{12}c_{13}  &  s_{12} c_{13}  &  s_{13} e^{-i\delta_{13}} \\
-s_{12} c_{23}-c_{12}s_{23}s_{13}e^{i\delta_{13}} &
c_{12} c_{23}- s_{12} s_{23} s_{13} e^{i\delta_{13}} &
s_{23} c_{13}  \\
s_{12} s_{23}-c_{12}c_{23}s_{13}e^{i\delta_{13}} &
-c_{12} s_{23}- s_{12} c_{23} s_{13} e^{i\delta_{13}} &
c_{23} c_{13}
\ea
\right] .
\ee
Here $c_{ij} \equiv \cos{\theta_{ij}}$ and
$s_{ij} \equiv \sin{\theta_{ij}}$, with $i$ and $j$ being
``generation'' labels ($i,j=1,2,3$).
The real angles $\theta_{12}$, $\theta_{23}$ and $\theta_{13}$
can all be made to lie in the first quadrant, by an appropriate
redefinition of quark field phases; then,
$c_{ij}\geq 0$, $s_{ij}\geq 0$ and $0\leq \delta_{13}\leq 2\pi$.

It is an empirical fact that the CKM matrix shows a
hierarchical pattern, with the diagonal elements being very close
to one, the ones connecting the two first generations having a size
$\lambda\simeq\sin{\theta_C}\approx 0.22$, the mixing between the
second and third families being of order $\lambda^2$, and
the mixing between the first and third quark flavours
having a much smaller size of about $\lambda^3$.
It is then quite practical to use the so-called Wolfenstein \cite{wo:2}
approximate representation of $V$:
\be\label{eq:wolfenstein}
{\bf V}\, =\,
\left[ \matrix{\displaystyle \ 1- {\lambda^2 \over 2} \hfill&
\displaystyle \ \ \ \ \ \ \lambda \hfill& \displaystyle \ \ \ \ \ A\lambda^
3(\rho  - i\eta) \hfill \cr\displaystyle \hfill& \displaystyle \hfill&
\displaystyle \hfill \cr\displaystyle \ \ \ -\lambda \hfill& \displaystyle \ \
\ \ \ 1 -{\lambda^ 2 \over 2} \hfill& \displaystyle \ \ \ \ \ A\lambda^ 2
\hfill
\cr\displaystyle \hfill& \displaystyle \hfill& \displaystyle \hfill
\cr\displaystyle \ A\lambda^ 3(1-\rho -i\eta) \hfill& \displaystyle \ \ \ \ \
-A\lambda^ 2 \hfill& \displaystyle \ \ \ \ \ \ \ \ \ 1 \hfill \cr} \right]\
+\ {\cal O}\left(\lambda^ 4 \right) \, .
\ee
The values of $A$, $\rho$ and $\eta$ are poorly
measured: $A=0.82\pm0.10$ and $|\rho|,|\eta|<0.5$.
CP is violated if $\eta\not=0$.

Since $\delta_{13}$ ($\eta$) is the oly possible source of CP violation,
the SM predictions for CP-violating phenomena are quite constrained.
Moreover, the CKM mechanism requires several necessary conditions
in order to generate
an observable CP-violation effect.
With only two fermion generations, the quark-mixing mechanism cannot
give rise to CP violation; therefore,
for CP violation to occur in a particular process,
all 3 generations are required to play an active role.
In the kaon system, for instance, CP-violation effects can only
appear at the one-loop level, where the top-quark is present.
In addition, all CKM-matrix elements must be non-zero and the quarks
of a given charge must be non-degenerate in mass. If any of these
conditions were not satisfied, the CKM-phase could be rotated away
by a redefinition of the quark fields. CP-violation effects
are then necessarily proportional to the product of all CKM-angles, and
should vanish in the limit where any two (equal-charge) quark-masses
are taken to be equal.
All these necessary conditions can be summarized in a very elegant
way as a single requirement on the original quark-mass matrices
${\bf m}$ and $\widetilde{\bf m}$ \cite{ref:jarlskog}:
\be
\mbox{\rm CP violation} \, \Longleftrightarrow \,
\mbox{\rm Im}\{\det[{\bf
mm^\dagger,\widetilde{m}\widetilde{m}^\dagger}]\}\not=0
\ee

Without performing any detailed calculation, one can make the
following general statements on the implications of the CKM mechanism
of CP violation:
\bi
\item 
Owing to unitarity, for any choice of $i,j,k,l$ (between 1 and 3),
\beqn\label{eq:J_relation}
\mbox{\rm Im}[V^{\phantom{*}}_{ij}V^*_{ik}V^{\phantom{*}}_{lk}V^*_{lj}]
\, =\, \cJ \sum_{m,n=1}^3 \epsilon_{ilm}\epsilon_{jkn}\, ,
\qquad\qquad\\
\cJ \, =\, c_{12} c_{23} c_{13}^2 s_{12} s_{23} s_{13} \sin{\delta_{13}}
\,\approx\, A^2\lambda^6\eta < 10^{-4}\, .
\eeqn
Any CP-violation observable involves the product
$\cJ$ \cite{ref:jarlskog}.
Thus, violations of the CP symmetry are necessarily small.
\item In order to have sizeable CP-violating asymmetries
[$(\Gamma - \overline{\Gamma})/(\Gamma + \overline{\Gamma})$], one should look
for very suppressed decays, where the decay widths already involve
small CKM matrix elements.
\item In the SM, CP violation is a low-energy phenomena in the sense that any
effect should dissapear when the quark-mass difference $m_c-m_u$ becomes
negligible.
\item $B$ decays are the optimal place for CP-violation signals to show up.
They involve small CKM elements and are the lowest-mass processes where
the 3 generations play a direct (tree level) role.
\ei

\section{Indirect and Direct CP Violation in the Kaon System}
\label{sec:mechanisms}

Any observable CP-violation effect is generated by the interference between
different amplitudes contributing to the same physical transition.
This interference can occur either through meson-antimeson mixing
or via final-state interactions, or by a combination of both effects.

\subsection{$K^0$-$\bar K^0$ Mixing}

The strangeness (flavour) quantum number is not conserved by
weak interactions.
Thus a $K^0$ state can be transformed into its antiparticle $\bar K^0$
(and analogously for $D^0$ and $B^0$ mesons).
Assuming CPT symmetry to hold,
the $2\times 2$ $K^0$-$\bar K^0$ mixing matrix
can be written as
\be\label{eq:mass_matrix}
{\cal M} =
\left(   \begin{array}{cc} M & M_{12} \\ M_{12}^* & M \ea   \right)
- {i\over 2}
\left(   \begin{array}{cc} \Gamma & \Gamma_{12} \\
          \Gamma_{12}^* & \Gamma \ea   \right) ,
\ee
where the diagonal elements $M$ and $\Gamma$ are real parameters.
If CP were conserved, $M_{12}$ and $\Gamma_{12}$ would also be real.
The physical eigenstates of ${\cal M}$ are
\be\label{eq:eigenstates}
| K_{S,L} \rangle \, = \, {1\over\sqrt{|p|^2 + |q|^2}} \,
       \left[ p \, | K^0 \rangle \, \mp\, q \, | \bar K^0 \rangle \right] ,
\ee
where
\be\label{eq:q/p}
{q\over p} \, \equiv \, {1 - \bar\varepsilon \over
         1 + \bar\varepsilon} \, = \,
  \left( {M_{12}^* - {i\over 2}\Gamma_{12}^* \over
          M_{12} - {i\over 2}\Gamma_{12}} \right)^{1/2} .
\ee
Clearly if $M_{12}$ and $\Gamma_{12}$ were real, then $q/p = 1$ and
$| K_{S,L} \rangle $ would correspond to the
CP-even
($K_1$) and CP-odd ($K_2$) states
$|K_{1,2}\rangle\equiv\left( |K^0\rangle\mp
|\bar K^0\rangle\right)/\sqrt{2}\, $
[we use the phase convention\footnote{
%
%
Since the flavour quantum number is conserved by strong interactions,
there is
some freedom in defining the phases of the flavour eigenstates.
In general, one
could use
$$
|K^0_\zeta\rangle \equiv e^{-i\zeta} |K^0\rangle , \qquad
|\bar K^0_\zeta\rangle \equiv e^{i\zeta} |\bar K^0\rangle ,
$$
which satisfy
$CP |K^0_\zeta\rangle = - e^{-2i\zeta} |\bar K^0_\zeta\rangle$.
Both basis are of course trivially related:
$M_{12}^\zeta = e^{2i\zeta} M_{12}$,
$\Gamma_{12}^\zeta = e^{2i\zeta} \Gamma_{12}$ and
$(q/p)_\zeta = e^{-2i\zeta} (q/p)$.
Thus, in general, $q/p\not=1$ does not necessarily imply CP violation.
CP is violated in the mixing matrix if $|q/p|\not=1$,
i.e. $\mbox{\rm Re}(\bar\varepsilon)\not=0$ and
$\langle K_L | K_S\rangle \not= 0$.
Note that
$\langle K_L | K_S\rangle_\zeta =\langle K_L | K_S\rangle$.
Another phase-convention independent quantity is
${q\over p}{\bar A_f\over A_f}$, where $A_f\equiv A(K^0\to f)$ and
$\bar A_f\equiv A(\bar K^0\to f)$, for any final state $f$.}
%
%
$CP |K^0\rangle = - |\bar K^0\rangle$].
Note that if the $K^0$-$\bar K^0$ mixing violates CP, the two mass
eigenstates are no longer orthogonal:
\be \langle K_L | K_S\rangle = {|p|^2-|q|^2 \over |p|^2+|q|^2}
\approx 2 \mbox{\rm Re}(\bar\varepsilon) .
\ee

The departure of $|p/q|$ from unity can be measured, by looking to
a CP-violating asymmetry in a flavour-specific decay,
i.e. a decay into a final state which can only be reached from an initial
$K^0$ (or $\bar K^0$) but not from both:
\be
K^0\to\pi^- l^+\nu_l , \qquad\qquad
\bar K^0\to\pi^+ l^-\bar\nu_l .
\ee
In the SM,
$|A(\bar K^0\to\pi^+ l^-\bar\nu_l)| = |A(K^0\to\pi^- l^+\nu_l)|$;
therefore,
\be
\delta \equiv
{\Gamma(K_L\to\pi^- l^+\nu_l) - \Gamma(K_L\to\pi^+ l^-\bar\nu_l)\over
\Gamma(K_L\to\pi^- l^+\nu_l) + \Gamma(K_L\to\pi^+ l^-\bar\nu_l)}
 = {|p|^2-|q|^2 \over |p|^2+|q|^2}
= {2 \mbox{\rm Re}(\bar\varepsilon)\over (1 + |\bar\varepsilon|^2)} .
\ee
The experimental measurement \cite{ref:PDG92},
$\delta = (3.27\pm 0.12)\times 10^{-3}$,
implies
\be\label{eq:Repsilon}
\mbox{\rm Re}(\bar\varepsilon) = (1.63\pm 0.06)\times 10^{-3} .
\ee

\subsection{Direct CP Violation}

If the flavour of the decaying meson $P$ is known,
any observed difference between the decay rate
$\Gamma(P\to f)$ and its CP conjugate $\Gamma(\bar P\to \bar f)$
would indicate that CP is directly violated in the decay amplitude.
One could study, for instance,
CP asymmetries in charged-kaon decays, such as $K^\pm\to\pi^\pm\pi^0$,
where the charge of the final pions clearly identifies the flavour
of the decaying kaon
(these types of decays are often referred to as self-tagging modes).
No positive signal has been reported up to date.

Since at least two interfering amplitudes are needed
to generate a CP-violating effect,
let us write the amplitudes
for the transitions $P \to f$ and
$\bar P \to \bar f$ as
\beqn
\label{eq:direct_b}
A[P \to f] & = & \, M_1 \, e^{i\phi_1}\, e^{i \alpha_1}\,
   +\, M_2 \, e^{i\phi_2}\, e^{i \alpha_2} \, ,
\\
A[\bar P \to \bar f] & = &
  M_1 e^{-i\phi_1} e^{i \alpha_1}\, +\, M_2 e^{-i\phi_2}e^{i \alpha_2} \, ,
\eeqn
where $\phi_1$, $\phi_2$ denote the weak phases, $\alpha_1$, $\alpha_2$
strong final-state phases (and/or strong phases between S- and P-wave
contributions in the case of baryon decays), and $M_1$, $M_2$ the moduli
of the matrix elements. The rate asymmetry is then given by
\be
\label{eq:direct_ratediff}
{\Gamma[P \to f] - \Gamma[\bar P \to \bar f] \over
\Gamma[P \to f] - \Gamma[\bar P \to \bar f]}
\,=\,
{-2 M_1 M_2 \sin{(\phi_1 - \phi_2)}
\sin{(\alpha_1 - \alpha_2)} \over
|M_1|^2 + |M_2|^2 + 2 M_1 M_2 \cos{(\phi_1 - \phi_2)}
\cos{(\alpha_1 - \alpha_2)}} .
\ee
Eq. (\ref{eq:direct_ratediff}) tells us that the following requirements
are needed in order to generate a direct-CP asymmetry:
\bi
\item Two (at least) interfering amplitudes.
\item Two different weak phases
[$\sin{(\phi_1 - \phi_2)}\not=0$].
\item Two different strong phases
[$\sin{(\alpha_1 - \alpha_2)}\not=0$].
\ei
Moreover, in order to get a sizeable asymmetry (rate difference / sum),
the two amplitudes $M_1$ and $M_2$ should be of comparable size.

In the kaon system, direct CP violation has been searched for in decays of
neutral kaons, where $K^0$-$\bar K^0$ mixing is also involved. Thus,
both direct and indirect CP-violation effects need to be taken into account,
simultaneously.
Since the $\pi^+\pi^-$ and $2\pi^0$ states are even
under CP, only the $K_1$ state could decay
into $2\pi$ if CP were conserved; $3\pi$'s at least would then
be required to
allow a hadronic decay of the $K_2$.
Therefore, owing to the phase-space suppression of the $K_L\to 3\pi$ decay
mode,
the $K_L\approx K_2 + \bar\varepsilon K_1$ state has
a much longer lifetime than
the $K_S\approx K_1 + \bar\varepsilon K_2$.
Since CP is violated, the $K_L$ does decay into $2\pi$.
The CP-violation signal is provided by the asymmetries:
\beqn\label{eq:etapm}
\eta_{+-} \equiv {A(K_L\to\pi^+\pi^-)\over A(K_S\to\pi^+\pi^-)}
  &\equiv & |\eta_{+-}| e^{i\phi_{+-}}
  \,\approx\, \varepsilon + {\varepsilon'\over 1 + \omega/\sqrt{2}} ,
\\ \label{eq:etazero}
\eta_{00} \,\equiv\, {A(K_L\to\pi^0\pi^0)\over A(K_S\to\pi^0\pi^0)}
  &\equiv & |\eta_{00}|\,  e^{i\phi_{00}}
  \,\approx\, \varepsilon - {2\varepsilon'\over 1 - \sqrt{2}\omega} ,\quad
\eeqn
where
\beqn
\varepsilon &\equiv & \bar\varepsilon + i \xi_0 , \\
\varepsilon' &\equiv &{i\over\sqrt{2}} \,\omega\, (\xi_2 - \xi_0) , \\
\omega &\equiv &{\mbox{\rm Re}(A_2)\over\mbox{\rm Re}(A_0)}\,
   e^{i(\delta_2-\delta_0)}.
\eeqn
$A_I$ and $\delta_I$ are the decay-amplitudes and strong phase-shifts
of isospin $I=0,2$ (these are the only two values allowed by Bose
symmetry for the final $2\pi$ state),
\be
A[K^0\to(2\pi)_I] \,\equiv\, i A_I e^{i\delta_I}, \qquad\qquad
A[\bar K^0\to(2\pi)_I] \,\equiv\, -i A_I^* e^{i\delta_I} ,
\ee
and
\be
\xi_I\,\equiv\, {\mbox{\rm Im}(A_I)\over\mbox{\rm Re}(A_I)} .
\ee
In Eqs. (\ref{eq:etapm}) and (\ref{eq:etazero}), terms quadratic in
the small CP-violating quantities have been neglected.

The parameter $\varepsilon$ is related to the indirect CP violation. Note that
$\varepsilon$ is a physical (measurable) phase-convention-independent
quantity, while $\bar\varepsilon$ is not [$\varepsilon$ =$\bar\varepsilon$
in the phase convention $\mbox{\rm Im}(A_0)=0$; however,
$\mbox{\rm Re}(\varepsilon) = \mbox{\rm Re}(\bar\varepsilon)$
in any convention].
Direct CP violation is measured through $\varepsilon'$, which is governed by
the phase-difference between the two isospin amplitudes.
The CP-conserving parameter $\omega$ gives the relative size between
these two amplitudes; experimentally, one finds a very big
enhancement of the $I=0$
channel with respect to the $I=2$ one,
which is known as the $\Delta I = 1/2$ rule:
\be
|\omega| \approx {1\over 22} , \qquad\qquad
\delta_2-\delta_0 = -45^\circ \pm 6^\circ .
\ee
The small size of $|\omega|$ implies a strong suppression of
$\varepsilon'$.

{}From the eigenvector equations for $K_S$ and $K_L$ one can easily
obtain the relation
\be
\bar\varepsilon \,\approx\, e^{i\phi_{SW}}\,
{\mbox{\rm Im}(M_{12}) - {i\over 2} \mbox{\rm Im}(\Gamma_{12})\over
\sqrt{\Delta M^2 + {1\over 4} \Delta\Gamma^2}} ,
\ee
where \cite{ref:PDG92}
$\Delta M \equiv M(K_L) - M(K_S) = (3.522\pm0.016)\times 10^{-12}$ MeV,
$\Delta\Gamma \equiv\Gamma(K_L)-\Gamma(K_S)\approx -\Gamma(K_S) =
-(7.377\pm 0.017)\times 10^{-12}$ MeV,
and
\be
\phi_{SW}\equiv\arctan{\left( {-2\Delta M\over\Delta\Gamma}\right)}
= 43.68^\circ\pm0.15^\circ
\ee
is the so-called superweak phase.
Since $\Delta\Gamma\approx -2\Delta M$, one has $\phi_{SW}\approx\pi/4$.
Moreover, $\Gamma_{12}$ is dominated by the
$K^0\to (2\pi)_{I=0}$ decay mode; therefore,
$\mbox{\rm Im}(\Gamma_{12})/ \mbox{\rm Re}(\Gamma_{12}) \approx -2\xi_0$.
Using these relations, one gets the approximate result
\be
\varepsilon\,\approx\, {e^{i\pi/4}\over\sqrt{2}} \,\left\{
{\mbox{\rm Im}(M_{12})\over 2 \,\mbox{\rm Re}(M_{12})} + \xi_0 \right\} .
\ee
Notice that $\delta_2-\delta_0 + \pi/2\approx \pi/4$, i.e.
\be
\varepsilon'\,\approx\, {e^{i\pi/4}\over\sqrt{2}}\,  |\omega | \,
(\xi_2 - \xi_0 ) .
\ee
Thus, owing to the particular numerical values of the neutral-kaon-decay
parameters, the phases of $\varepsilon$ and $\varepsilon'$ are nearly
equal.

The experimental world-averages quoted by the Particle Data Group
\cite{ref:PDG92} are
\beqn
|\eta_{+-}| &=& (2.268\pm0.023)\times 10^{-3} , \\
|\eta_{00}| &=& (2.253\pm0.024)\times 10^{-3} .
\eeqn
These two numbers are equal within errors, showing that indeed
$|\varepsilon'|<<|\varepsilon|$ as expected from the
$|\omega|$ suppression. Moreover,
from Eq. (\ref{eq:Repsilon}) and $\arg{(\varepsilon)}\approx\pi/4$,
we have $|\varepsilon|\approx 2.3\times 10^{-3}$, in good agreement
with the value extracted from $K^0\to 2\pi$.

The ratio $\varepsilon'/\varepsilon$
can be determined through the relation
\be
\mbox{\rm Re}\left({\varepsilon'\over\varepsilon}\right) \approx
{1\over 6} \left\{ 1 - \left| {\eta_{00}\over\eta_{+-}}\right|^2\right\} .
\ee
Two different experiments \cite{ref:NA31_ep,ref:E731_ep}
have recently reported a measurement of this
quantity:
\be
\mbox{\rm Re}\left({\varepsilon'\over\varepsilon}\right) \, = \,
\left\{
\begin{array}{rc}
(23.0\pm 6.5)\times 10^{-4}  & \qquad \mbox{\rm NA31}
   \mbox{\cite{ref:NA31_ep}} , \\
(7.4\pm5.9)\times 10^{-4} & \qquad \mbox{\rm E731}
  \mbox{\cite{ref:E731_ep}} .
\ea \right.
\ee
The NA31 measurement provides
evidence for a non-zero value of $\varepsilon'/\varepsilon$ (i.e.
direct CP violation), with a statistical significance of more than three
standard deviations. However, this is not supported by
the E731 result, which is compatible with
$\varepsilon'/\varepsilon = 0$, thus with no direct CP violation.
The probability for the two results being statistically compatible is
only 7.6\%.

Clearly, new experiments with a better sensitivity are required in order
to resolve this discrepancy.
A next generation of $\varepsilon'/\varepsilon$ experiments is already
under construction at CERN \cite{ref:cern} and Fermilab \cite{ref:fnal}.
Moreover, a dedicated $\phi$ factory (DA$\Phi$NE), providing
large amounts of tagged $K_S$, $K_L$ and $K^\pm$
($\phi\to K\bar K$), is already being built
at Frascati \cite{ref:DAPHNE}. The goal of all these experiments is to
reach sensitivities better than $10^{-4}$.
In the meantime,
a much modest $10^{-3}$ sensitivity
should be reached by the CPLEAR experiment \cite{ref:nakada},
presently running at CERN.

\subsection{Time Evolution}

$K^0$-$\bar K^0$ mixing implies that a state which was originally produced
as a $K^0$ or a $\bar K^0$ will not develop in time in a purely
exponential fashion.
The time-dependent amplitudes for the decay into a given final state $f$
are given by:
\beqn
A(K^0\to f) &\sim &
\left\{ A(K_S\to f)\, e^{-iM_St}\, e^{-\Gamma_S t/2} +
A(K_L\to f)\, e^{-iM_Lt}\, e^{-\Gamma_L t/2} \right\} , \quad\\
A(\bar K^0\to f) &\sim &
\left\{ A(K_S\to f)\, e^{-iM_St}\, e^{-\Gamma_S t/2} -
A(K_L\to f)\, e^{-iM_Lt}\, e^{-\Gamma_L t/2} \right\} . \quad
\eeqn
In terms of the ratio
\be
\eta_f \equiv {A(K_L\to f)\over A(K_S\to f)} \equiv |\eta_f| e^{i\phi_f} ,
\ee
the time evolution of the decay rates can then be written as
\beqn
\Gamma(K^0\to f) &\sim &e^{-\Gamma_S t} + |\eta_f|^2\, e^{-\Gamma_L t}
  + 2 |\eta_f| \,
\cos{\left(\phi_f - \Delta M t\right)}\, e^{-(\Gamma_L+\Gamma_S)t/2} ,
\\
\Gamma(\bar K^0\to f) &\sim &e^{-\Gamma_S t} + |\eta_f|^2\, e^{-\Gamma_L t}
  - 2 |\eta_f| \,
\cos{\left(\phi_f - \Delta M t\right)}\, e^{-(\Gamma_L+\Gamma_S)t/2} .
\eeqn

By measuring the decay rate as a function of time,
the ratio $\eta_f$ (both modulus and phase)  and the mass-difference
$\Delta M$ can be obtained.
For the dominant $2\pi$ modes, the measured phases \cite{ref:PDG92},
\beqn
\phi_{+-} &=& (46.6\pm1.2)^\circ , \\
\phi_{00} &=& (46.6\pm2.0)^\circ ,
\eeqn
are very close to $\pi/4$, as expected.

\subsection{SM Predictions}
\label{subsec:predictions}

The CKM mechanism generates CP-violation effects both in the
$\Delta S=2$ $K^0$-$\bar K^0$ transition (box-diagrams) and in the
$\Delta S=1$ decay amplitudes (penguin diagrams).
Although a straightforward and well-defined technique,
which makes use of the Operator Product Expansion,
is available for a short-distance analysis of these
interactions, the final quantitative predictions are obscured
by the presence of hadronic matrix-elements of weak four-quark
operators, which are governed by long-distance physics.

\begin{figure}[htb]
\vfill
\caption{$\Delta S=2$ box diagrams.}
\label{fig:box_diagram}
\end{figure}

\begin{figure}[htb]
\caption{$\Delta S=1$ penguin diagrams.}
\label{fig:penguin}
\end{figure}

Only one such operator appears in the $K^0$-$\bar K^0$ mixing
analysis.
Including the short-distance QCD corrections, the box-diagram
calculation of $M_{12}$ yields
\beqn\label{eq:box}
M_{12} & = &{G_F^2 M_W^2\over 16\pi^2} \left\{
\lambda_c^2 \,\eta_1\, S(r_c) + \lambda_t^2\, \eta_2\, S(r_t) +
2 \lambda_c\lambda_t\, \eta_3\, S(r_c,r_t)\right\}
\quad\nonumber\\
&& \times\, {1\over 2 M_K} \, \alpha_s(\mu^2)^{-2/9}\,
\langle\bar K^0|\left(\bar s\gamma^\mu(1-\gamma_5)d\right)
\left(\bar s\gamma_\mu(1-\gamma_5)d\right)|K^0\rangle \, ,
\eeqn
where
\be
\lambda_i \,\equiv\, V^{\phantom{*}}_{is}\, V_{id}^*
\qquad\qquad (i=u,c,t) ,
\ee
and $S(r_i)$, $S(r_i,r_j)$ are functions of $r_i\equiv (m_i/M_W)^2$.
Owing to the unitarity of the CKM matrix,
$\lambda_u + \lambda_c + \lambda_t = 0$, and the contributions of the
up, charm and top quarks to the box diagram add to zero in the limit
of massless quarks (GIM mechanism \cite{ref:GIM}).
The loop functions $S(r_i)$ and $S(r_i,r_j)$ are then very sensitive
to the quark masses [$S(r_i)\approx r_i$, for $r_i<<1$].
For large $m_t$ the second term in Eq. (\ref{eq:box}) dominates.

The factors $\eta_i$ represent short-distance
QCD corrections to the lowest-order box-diagram calculation
($\eta_i = 1$ in the absence of QCD effects):
$\eta_1\approx 0.85$, $\eta_2\approx 0.57$ and $\eta_3\approx 0.36$
\cite{ref:bh92}.
In addition, one needs to compute the hadronic matrix element of
the $\Delta S=2$ four-quark operator in Eq. (\ref{eq:box}), which is
usually parametrized in terms of the so-called $B_K$ parameter:
\be\label{eq:bk}
\alpha_s(\mu^2)^{-2/9}\,
\langle\bar K^0|\left(\bar s\gamma^\mu(1-\gamma_5)d\right)
\left(\bar s\gamma_\mu(1-\gamma_5)d\right)|K^0\rangle \,
\equiv\, 2 \left(1+{1\over 3}\right) \left( \sqrt{2} f_K M_K\right)^2
B_K .
\ee
$B_K=1$ corresponds to the factorization approximation, which
consists in splitting the matrix element in a product of two
currents, by
inserting the vacuum in all possible ways
[$\langle\bar K^0|\left(\bar s\gamma^\mu(1-\gamma_5)d\right)|\emptyset
\rangle\,\langle\emptyset|
\left(\bar s\gamma_\mu(1-\gamma_5)d\right)|K^0\rangle$].
Clearly, this approximation can only be taken as an
order-of-magnitude estimate, since it completely ignores
the renormalization-group factor $\alpha_s(\mu^2)^{-2/9}$,
where $\mu$ is an arbitrary renormalization scale.
The total product (and therefore $B_K$) in Eq. (\ref{eq:bk})
is of course $\mu$-independent, because the dependence on the
renormalization scale is exactelly cancelled by the hadronic
matrix element. Unfortunately, it is very difficult to make
a calculation of this matrix element from first principles.
Table \ref{tab:bk} shows the values of $B_K$ obtained by
various methods.
The present uncertainty associated with the size of the
hadronic matrix element
is reflected in the broad range of calculated $B_K$ values.

\begin{table}
\begin{center}
\begin{tabular}{|c|c|c|}
\hline $B_K$ & Method & Reference \\ \hline
$0.33$ & Lowest-order Chiral Perturbation Theory &
\protect\cite{ref:dgh82}\\
$0.39\pm0.10$ & QCD Sum Rules + Chiral Symmetry &
\protect\cite{ref:pr85} \\
$0.5\pm0.1\pm0.2$ & QCD Sum Rules (3-point function) &
\protect\cite{ref:bdg88}\\
$0.4\pm0.2$ & Effective Action & \protect\cite{ref:pr91a}\\
$0.38^{+0.10}_{-0.02}$ & Estimate of $\cO(p^4)$ Chiral corrections &
\protect\cite{ref:bruno}\\
$3/4$ & Leading $1/N_c$ & \\
$0.70\pm0.10$ & $1/N_c$ expansion & \protect\cite{ref:bbg88}\\
$ 0.8\pm 0.2$ & Lattice & \protect\cite{ref:lattice} \\ \hline
\end{tabular}
\end{center}
\caption{Values of $B_K$ obtained by various methods.}
\label{tab:bk}
\end{table}

In order to compute $\varepsilon$,
$\mbox{\rm Im}(M_{12})$ is obtained from Eq. (\ref{eq:box})
while $\mbox{\rm Re}(M_{12})$, which is much more sensitive to
long-distance effects, is taken from the measured neutral-kaon
mass diference $\Delta M$.
The result depends on the unknown top mass and on the values of the
CKM elements. Using the Wolfenstein parametrization (\ref{eq:wolfenstein}),
the experimental value of $\varepsilon$ specifies  a
hyperbola in the $(\rho,\eta)$ plane \cite{ref:bh92}:
\be\label{eq:hyperbola}
\eta \left[ (1-\rho) A^2 r_t^{0.76} + P_C \right] A^2 B_K = 0.50 ,
\ee
where $P_C\approx 2/3$ contains the $cc$ and $tc$ box-diagram
contributions\footnote{
The power-like dependence on $r_t$ (and similar ones that will appear in
the following sections) represents a numerical fit to the exact loop
functions \protect\cite{ref:bh92}. In the range 100 GeV $< m_t <$
200 GeV, the exact result is reproduced to an accuracy better than 3\%.}.

The theoretical estimate of $\varepsilon'/\varepsilon$
is much more involved, because ten
four-quark operators need to be considered
in the analysis and the presence of cancellations between different
contributions tends to amplify the sensitivity to the not
very well-controlled long-distance effects.
A detailed discussion has been given in refs. \cite{ref:ep}.
For large values of the top-mass, the $Z^0$-penguin contributions
strongly suppress the expected value of $\varepsilon'/\varepsilon$,
making the final result very sensitive to $m_t$.
In the presently favoured range of top masses,
$m_t \sim 100-200 \ GeV$, the theoretical estimates \cite{ref:ep} give
$\varepsilon'/\varepsilon \sim (2 - 27)\times 10^{-4}$,
with large uncertainties.
More theoretical work is needed in order to get firm predictions.

\section{Strong CP Violation}
\label{sec:strong_CP}

CP violation could also originate from and additional term in the
QCD Lagrangian,
\be\label{eq:strong}
\cL_\theta = \theta_0 {g^2_s\over 64\pi^2} \epsilon^{\mu\nu\rho\sigma}
 \sum_a G^{(a)}_{\mu\nu} G^{(a)}_{\rho\sigma} \, ,
\ee
which violates P, T and CP.
Although (\ref{eq:strong}) is a total derivative, it can give rise
to observable effects because of the non-perturbative structure
of the QCD vacuum (see ref. \cite{ref:pr91b} for a detailed discussion).

One could try to impose P and T conservation in strong interactions,
i.e. $\theta_0 = 0$. However, owing to the axial anomaly of the
$U(1)_A$ current,
a non-zero value of $\theta_0$
would be again generated when diagonalizing the quark-mass matrices
${\bf m}$ and $\widetilde{\bf m}$.
A $U(1)_A$ rotation is needed to eliminate a global phase of the
quark-mass matrices, but due to the existence of a quantum anomaly,
the full theory is not invariant under this transformation; the
phase can be shifted from the quark-mass matrices to $\cL_\theta$,
but it cannot be eliminated.
In fact, the physical parameter (i.e. the one which
remains invariant under the phase rotation) is not quite
$\theta_0$, but rather the combination
\be
\theta\equiv\theta_0 + \arg{\left\{ \det{ \left(
  {\bf m}\right) } \det{ \left( \widetilde{\bf m} \right) }\right\} } .
\ee

A non-zero value of $\theta$ could lead to observable effects in
flavour-conserving transitions.
It may generate, in particular, a sizeable neutron electric dipole
moment, which very refined experiments have constrained down to a very
high precision \cite{ref:edm}:
\be
d_n^\gamma < 12\times 10^{-26} \, e \,\mbox{\rm cm}
\quad\mbox{\rm (95\% C.L.)} .
\ee
This provides a stringent upper limit on $|\theta|$.
The more recent estimate of $d_n^\gamma$ \cite{ref:pr91b}, done
in the framework of Chiral Perturbation Theory (ChPT), gives
\be
|\theta| < 5\times 10^{-10} \, .
\ee

  The smallness of this number, makes the $\theta$ effect completely
irrelevant for the phenomenology of CP violation in weak transitions.
However, it leaves as an open question the reason for such a small
quantity. The initial value of $\theta_0$ and the messy phases present in the
original Yukawa couplings should conspire to generate a huge cancellation
giving rise to such a tiny value of $\theta$!
This is usually known as the ``strong CP problem''.
The most plausible explanation is that the effective $\theta$ is probably
just zero, because there is some additional symmetry which makes it
unobservable (i.e. it can be finally rotated away anyhow)
\cite{ref:pq}.
Unfortunately, a clear solution of the problem is still missing.

\section{Rare K Decays}
\label{sec:rare}

High precision experiments on rare kaon decays \cite{ref:k_rev}
offer the
exciting possibility of unravelling new physics beyond
the SM. Searching for forbidden
flavour-changing processes ($K_L\to\mu e$, $K_L\to\pi^0\mu e$,
$K^+\to\pi^+\mu e$, ...)
at the $10^{-10}$ level, one is actually exploring
energy-scales above the 10 TeV region. The study of allowed (but highly
suppressed) decay modes provides,
at the same time, very interesting tests of
the SM itself. Electromagnetic-induced non-leptonic weak
transitions and higher-order weak processes are a useful tool to improve
our understanding of the interplay among electromagnetic, weak and strong
interactions. In addition, new signals of CP violation, which would help
to elucidate the source of CP-violating phenomena, can be looked for.

\subsection{$K_L\to\pi^0\nu\bar\nu$}

Long-distance effects play a negligible role in the decays
$K\to\pi\nu\bar\nu$, which proceed through $W$-box and $Z$-penguin
diagrams. The resulting amplitudes are proportional to the matrix
element of the $\Delta S=1$ vector current,
\be
T(K^+\to\pi^+\nu\bar\nu) \sim \sum_{i=u,c,t} F(\lambda_i,r_i)\,
(\bar\nu_L\gamma_\mu\nu_L) \,
\langle \pi|(\bar s \gamma^\mu d)|K\rangle \, ,
\ee
which is known from the $K_{l3}$ decays.

The decay $K^+\to\pi^+\nu\bar\nu$
provides then a good test of the radiative structure of the SM,
and could be used to extract clean information on the CKM factors.
Summing over the three neutrino flavours,
its branching ratio is expected to be around $(1-5)\times 10^{-11}$
\cite{ref:bh92}, while the present experimental upper bound is
$Br(K^+\to\pi^+\nu\bar\nu)<5.2\times 10^{-9}$ (90\% C.L.) \cite{ref:k_rev}.
Experiments aiming to reach a sensitivity at the level of the SM
prediction are already under way.

The CP-violating decay $K_L\to\pi^0\nu\bar\nu$ has been suggested
\cite{littenberg} as a good candidate to look for (nearly) pure
direct CP-violating transitions.
Its decay amplitude its related to the $K^+\to\pi^+\nu\bar\nu$ one
by isospin:
\be
T(K_L\to\pi^0\nu\bar\nu) = {1\over 2} \left\{
(1+\bar\varepsilon) T(K^+\to\pi^+\nu\bar\nu) -
(1-\bar\varepsilon) T(K^-\to\pi^-\nu\bar\nu) \right\} \, .
\ee
The contribution coming from indirect
CP-violation via $K^0$-$\bar K^0$ mixing is predicted to be around
$10^{-15}$ \cite{littenberg}.
Direct CP-violation generates a much bigger contribution \cite{ref:bh92}:
\be
Br(K_L\to\pi^0\nu\bar\nu) \,\approx\, 0.82\times 10^{-10} r_t^{1.18}
A^4 \eta^2 \, .
\ee
The clean observation of just a single
unambiguous event would indicate the existence of CP-violating
$\Delta S = 1$ transitions.
The possibility of detecting such a decay mode is,
of course, a big experimental challenge.
The present (90\% C.L.) experimental limit is
$Br(K_L\to\pi^0\nu\bar\nu)<2.2\times 10^{-4}$.
A much better sensitivity of about $10^{-7}$ is expected
to be achieved in the
near future \cite{ref:k_rev}.

\subsection{$K_L \rightarrow \pi^0 e^+ e^-$}

 The $K_L\to\pi^0 e^+ e^-$ decay looks more promising.
If CP were an exact symmetry,
only the CP-even state $K_1$ could decay via one-photon emission, while
the decay of the CP-odd state $K_2$ would proceed through a two-photon
intermediate state and, therefore, its decay amplitude would be suppressed
by an additional power of $\alpha$. When CP violation is taken into account,
however, an $\cO(\alpha)$ $K_L \rightarrow \pi^0 e^+ e^-$ decay
amplitude is induced, both through the small $K_1$ component of the $K_L$
($\varepsilon$ effect) and through direct CP violation in the
$K_2 \rightarrow \pi^0 e^+ e^-$ transition. The electromagnetic suppression
of the CP-conserving amplitude then makes it plausible that this decay is
dominated by the CP-violation contributions.

 The branching ratio induced by the direct CP-violation amplitude
 is predicted \cite{ref:ddgb}
to be around $10^{-11}$,
the exact number depending on the values of $m_t$ and the quark-mixing
angles \cite{ref:bh92}:
\be
Br(K_L\to\pi^0 e^+ e^-)\Big|_{\mbox{\rms Direct CP}} \approx
0.23\times 10^{-10} r_t^{1.18} A^4 \eta^2 \, .
\ee

The indirect CP-violating contribution
is given by the $K_S \rightarrow \pi^0 e^+ e^-$ amplitude
times the CP-mixing parameter $\varepsilon$.
Using ChPT techniques, it is possible to
relate \cite{ref:epr88} the $K_S$ decay amplitude to the measured
$K^+\to\pi^+e^+e^-$ transition. Present data implies \cite{ref:mexico}
\be
Br(K_L \rightarrow \pi^0 e^+ e^-)\Big|_{\mbox{\rms Indirect CP}} \le
           1.6 \times 10^{-12}.
\ee
Therefore, the interesting direct
CP-violating contribution is expected to be bigger than the
indirect one. This is very different from the situation in
$K \rightarrow \pi \pi$, where the contribution due to mixing
completely dominates.

   The present experimental upper bound \cite{ref:k_rev}
(90\% C.L.)
\be
Br(K_L \rightarrow \pi^0 e^+ e^-)\Big|_{\mbox{\rms Exp}}
 < 5.5 \times 10^{-9},
\ee
is still far away from the expected SM signal,
but the prospects
for getting the needed sensitivity of around $10^{-12}$ in
the next few years are rather encouraging.
In order to be able to interpret a future experimental measurement of
this decay as a CP-violating signature, it is first necessary, however,
to pin down the actual
size of the two-photon-exchange CP-conserving amplitude.

The $K_L\to\pi^0\gamma\gamma$ amplitude can be computed within
ChPT \cite{ref:epr87}.
One can then estimate the two-photon-exchange contribution
to $K_L\to\pi^0e^+e^-$,
by taking the absorptive part due to the two-photon discontinuity as an
educated guess of the actual size of the complete amplitude
\cite{ref:epr88,ref:epr90}.
At the lowest non-trivial order in the momentum expansion, $\cO(p^4)$,
the $K_L\to\pi^0e^+e^-$ decay
amplitude is
strongly suppressed (it is proportional to $m_e$), owing to the
helicity structure of the
$K_L\to\pi^0\gamma\gamma$ decay amplitude
\cite{ref:epr88}:
\be
Br(K_L \rightarrow \pi^0 \gamma ^* \gamma ^* \rightarrow \pi^0
     e^+ e^-)\Big|_{\cO(p^4)} \,\sim\, 5 \times 10^{-15} .
\ee
This helicity suppression is, however, no longer true at the next order
in the chiral expansion, because an additional Lorentz structure
is then allowed in the decay amplitude.
An estimate of the dominant
$\cO(p^6)$ contribution can be done, by using the
measured \cite{ref:NA31_92}
photon spectrum in the decay $K_L\to\pi^0\gamma\gamma$.
The most recent analysis
\cite{ref:CEP93} gives
\be
Br(K_L \rightarrow \pi^0 \gamma^* \gamma^* \rightarrow \pi^0 e^+ e^-)
 \Big|_{\cO(p^6 )} \,\sim\,
(0.3-1.8) \times 10^{-12} \, ,
\ee
 implying that this decay is in fact dominated by the CP-violation
contribution.

\subsection{Longitudinal Muon Polarization in $K_L\to\mu^+\mu^-$}

The longitudinal muon polarization $\cP_L$
in the decay $K_L\to\mu^+\mu^-$ is an interesting measure of CP violation.
As for every CP-violating observable in the neutral kaon system,
both indirect and direct CP-violation contributions need to be
considered.

   In the SM, the direct CP-violating amplitude is
induced by Higgs exchange with an effective one-loop flavour-changing
$\bar s d H$ coupling \cite{ref:BL86}.
The present lower bound \cite{ref:HIGGS}
on the Higgs mass $m_H>63.5$ GeV ($95 \%$ C.L.), implies
\cite{ref:BL86,ref:GN89} a
conservative upper limit
$|\cP_{L,\mbox{\rms Direct}}| < 10^{-4}$.
A much larger value $\cP_L \sim O(10^{-2})$ appears quite naturally
in various extensions of the SM \cite{ref:MO93}.
It is worth emphasizing that $\cP_L$ is especially
sensitive to the presence of light scalars with CP-violating
Yukawa couplings. Thus, $\cP_L$ seems to be a good signature to look
for new physics beyond the SM; for this to be the case,
however, it is very important to have a good quantitative
understanding of the SM prediction to allow us to infer,
from a measurement of $\cP_L$, the existence of a new CP-violation
mechanism.

The $K_1^0\to\mu^+\mu^-$ decay amplitude can be unambiguously
calculated in ChPT \cite{ref:EP91}. This
allows us to make a reliable estimate\footnote{
Taking only the absorptive parts of the $K_{1,2}\to\mu^+\mu^-$
amplitudes into account,
a value $|\cP_{L,\varepsilon}| \approx 7\times 10^{-4}$ was
estimated previously \protect\cite{ref:HE83}.
However, this is only one out of four contributions
to $\cP_L$ \protect\cite{ref:EP91},
which
could all interfere constructively with unknown magnitudes.
}
of the contribution to $\cP_L$ due to $K^0$-$\bar K^0$ mixing
\cite{ref:EP91}:
\be\label{eq:p_l}
1.9 < |\cP_{L,\varepsilon}| \times 10^3 \Biggl( {2 \times 10^{-6} \over
Br(K_S\to\gamma\gamma)} \Biggr)^{1/2} < 2.5  .
\ee
Taking into account
the present experimental errors \cite{ref:NA31_87}
in $Br(K_S\to\gamma\gamma)$
[$Br = (2.4\pm1.2)\times 10^{-6}$]
and
the inherent theoretical uncertainties due to uncalculated
higher-order corrections,
one can conclude that experimental indications for
$|\cP_L|>5\times 10^{-3}$ would constitute clear evidence
for additional
mechanisms of CP violation beyond the SM.

\section{B decays}
\label{sec:bottom}

Differences of rates that signal CP violation
are proportional to the small product $A^2\lambda^6\eta$, but the
corresponding asymmetries (difference / sum)
are enhanced in B-decay relative to K-decay because the
B-decay widths involve much smaller CKM elements
($|V_{cb}|^2$ or $|V_{ub}|^2 \,<<\, |V_{us}|^2$).
If the SM is correct, sizeable CP-violation asymmetries should be expected to
show up in many decay modes of beauty particles \cite{ref:ecfa}.

\subsection{Indirect CP Violation}

The general formalism to describe mixing among the neutral
$B^0$ and $\bar B^0$
mesons is completely analogous to the one used in the kaon sector.
However,
the physical mass-eigenstates [$CP |B^0\rangle = - |\bar B^0\rangle $]
\be\label{eq:B_eigenstates}
| B_\mp \rangle \, = \, {1\over\sqrt{|p|^2 + |q|^2}} \,
       \left[ p \, | B^0 \rangle \, \mp\, q \, | \bar B^0 \rangle \right]
\ee
have now a comparable lifetime, because many decay modes are common to both
states and therefore the available phase space is similar.

The flavour-specific decays
\be
B^0\to l^+\nu_l X \, , \qquad\qquad\qquad \bar B^0\to l^-\bar\nu_l X \, ,
\ee
provide the most direct way to measure the amount of CP violation in
the $B^0$-$\bar B^0$ mixing matrix.
The asymmetry between the number of $l^+l^+$ and $l^-l^-$ pairs produced
in the processes $e^+e^-\to B^0\bar B^0\to l^\pm l^\pm X$
is easily found to be
\be
\label{eq:a_SL_def}
a_{SL} \equiv {N(l^+l^+) - N(l^-l^-) \over N(l^+l^+) + N(l^-l^-)}
= {\left|p/q\right|^2 - \left|q/p\right|^2 \over
   \left|p/q\right|^2 + \left|q/p\right|^2}
\approx 4 \mbox{\rm Re}(\bar\varepsilon_B)  .
\ee

Unfortunately, this $\Delta B = 2$ asymmetry is expected to be quite tiny
in the SM, because
$|\Delta\Gamma/\Delta M| \approx |\Gamma_{12}/M_{12}| << 1 \,\,$
[$\Delta M \equiv M_{B_+}-M_{B_-}$,
$\Delta\Gamma\equiv\Gamma_{B_+}-\Gamma_{B_-}$].
This can be easily understood, by looking to the relevant box diagrams
contributing to the  $B^0$-$\bar B^0$ transition;
the mass mixing is dominated
by the top-quark graph, while the decay  amplitudes get obviously its main
contribution from the $b\to c$ transition. Thus,
\be\label{eq:il_fuction}
{\Gamma_{12}\over M_{12}}\,\approx\, {3\pi\over 2}\, {m_b^2\over m_t^2}
\, {1\over E^\prime(r_t)}\, << \, 1 ,
\ee
where \cite{il} $E^\prime(r_t)$ is a slowly decreasing function of
$r_t\,\,$ [$E'(0)=1$, 
$E'(\infty)=1/4$].
One has then [see Eq. (\ref{eq:q/p})]
\be
\left| {q\over p}\right| \approx 1 + {1\over 2}
\left| {\Gamma_{12}\over M_{12}}\right| \sin{\phi_{\Delta B=2}} ,
\ee
where
\be
\phi_{\Delta B=2} \equiv \arg{\left({M_{12}\over\Gamma_{12}}\right)} .
\ee
The factor $\sin{\phi_{\Delta B=2}}$ involves an additional GIM suppression,
\be
\sin{\phi_{\Delta B=2}} \,\approx\, {8\over 3} {m_c^2-m_u^2\over m_b^2}
\,\mbox{\rm Im}\left({V^{\phantom{*}}_{cb} V_{cq}^*\over
V^{\phantom{*}}_{tb} V_{tq}^*}\right) ,
\ee
implying a value of $|q/p|$ very close to 1. Here, $q\equiv d,s$
denote the
corresponding CKM matrix elements for $B^0_q$ mesons.
Therefore, one expects
\be
\label{eq:a_SL_expected}
a_{SL} \leq \left\{ \ba 10^{-3}  \qquad (B^0_d), \\
10^{-4}  \qquad (B^0_s). \ea
\right.\ee
The observation of an asymmetry $a_{SL}$ at the percent level,
would then be
a clear indication of new physics beyond the SM.

\subsection{Direct CP violation}
\label{subsec:direct}

Direct CP violation could be established by measuring a non-zero rate
asymmetry in $B^\pm$ decays.
One example is the decay $B^\pm \to K^\pm \rho^0$ which proceeds via
a tree- and a penguin-diagram
(Fig.\nobreak\ \ref{fig:direct_cp_diagrams}),
the weak couplings of which are given
by $V^{\phantom{*}}_{ub} V_{us}^*\approx A\lambda^4(\rho-i\eta)$
and $V^{\phantom{*}}_{tb} V_{ts}^*\approx -A\lambda^2$,
respectively\footnote{
Since $m_u, m_c << M_W$, we can neglect the small quark-mass corrections
in the up and charm penguin contributions. These two diagrams
then differ in their CKM factors only, and their sum is regulated by the
same CKM factor than the top-quark loop, due to the unitarity
of the CKM matrix.}.
Although the penguin contribution is of higher-order in the strong
coupling, and suppressed by the loop factor $1/(16\pi^2)$,
one could expect both amplitudes to be of comparable size, owing
to the additional $\lambda^2$ suppression factor of the tree diagram.
The needed strong-phase difference can be generated through the
absorptive part of the penguin diagram, corresponding to
on-shell intermediate particle rescattering \cite{BSS}.
Therefore, one could expect a sizeable asymmetry, provided the
strong-phase difference is not too small.
However, a very large number of $B^{\pm}$ is required, because
the branching ratio is quite suppressed ($\sim 10^{-5}$).
Other decay modes such as $B^\pm\to K^\pm K_S,K^\pm K^{*0}$ \cite{GH} involve
the interference between penguin diagrams only and might show
sizeable CP-violating asymmetries as well, but the corresponding
branching fractions are expected to be even smaller than the previously
discussed one.

\begin{figure}[htb]
\vfill
\caption{Feynman diagrams contributing to $B^- \to K^- \rho^0$}
\label{fig:direct_cp_diagrams}
\vfill
\end{figure}

The two interfering amplitudes can also be generated through
other mechanisms. For instance, one can have an interplay between two
different cascade processes \cite{CS,BS} like
$B^-\to D^0 X^-\to K_S Y X^-$
and  $B^-\to\bar D^0 X^-\to K_S Y X^-$.
Another possibility would be an interference between
two  tree-diagrams corresponding to two different decay mechanisms like
direct decay (spectator) and weak annihilation \cite{BJ81}.
Direct CP violation could also be studied in decays of bottom
baryons \cite{Baryon},
where it could show up as a rate asymmetry and in various decay
parameters.

Note that, for all these flavour-specific decays,
the necessary presence of strong phases makes very difficult to extract
useful information on the CKM  factors from their measured
CP asymmetries.
Nevertheless, the experimental observation of a non-zero  CP-violating
asymmetry in any of these decay modes would be a major milestone in our
understanding of CP-violation phenomena, as it would clearly establish
the existence of direct CP violation in the decay amplitudes.

\subsection{Interplay Between Mixing and Direct CP Violation}
\label{subsec:interplay}

There are quite a few non-leptonic final states which are reachable
both from a $B^0$ and a $\bar B^0$. For these flavour non-specific decays
the $B^0$ (or $\bar B^0$) can decay directly to the given final state $f$,
or do it after the meson has been changed to its antiparticle via the
mixing process; i.e. there are two different amplitudes,
$A(B^0\to f)$ and $A(B^0\to\bar B^0\to f)$, corresponding to two possible
decay paths. CP-violating effects can then result from the interference
of these two contributions.

The time evolution of a state which was originally produced
as a $B^0$ or a  $\bar B^0$  is given by
\be\label{eq:evolution}
\left( \ba | B^0(t) \rangle  \\ | \bar B^0(t) \rangle \ea \right)
 =
\left( \begin{array}{cc} g_1(t)  & {q\over p} g_2(t) \\
     {p\over q} g_2(t) & g_1(t) \ea \right)
\left( \ba | B^0 \rangle  \\ | \bar B^0 \rangle \ea \right) .
\ee
where
\be\label{eq:g}
\left( \ba g_1(t) \\ g_2(t) \ea \right) =
e^{-iMt} e^{-\Gamma t/2}
\left( \ba \cos{[(\Delta M - {i\over 2} \Delta\Gamma) t/2]} \\
   -i \sin{[(\Delta M - {i\over 2} \Delta\Gamma) t/2]} \ea \right) .
\ee
Since for $B^0$ mesons $|\Delta\Gamma/\Delta M| <<1$, we will neglect the
tiny $\Delta\Gamma$ corrections in what follows.

  The time-dependent decay probabilities for the decay of a neutral
$B$ meson created at the time $t_0=0$ as a pure $B^0$
($\bar B^0$) into the final state $f$ ($\bar f\equiv CP\, f$) are
\beqn
\label{eq:decay_b}
  \Gamma[B^0(t)\to f] & \propto & {1\over 2} e^{-\Gamma t} |A_f|^2
        \left\{
[1 + |\bar\rho_f|^2] +  [1 - |\bar\rho_f|^2] \cos{(\Delta M t)}
\right. \quad\\ \nonumber & & \qquad\qquad\quad\left.
  - 2 \mbox{\rm Im}\left( {q\over p} \bar\rho_f\right) \sin{(\Delta M t)}
    \right\} ,
    \\
\label{eq:decay_bbar}
 \Gamma[\bar B^0(t)\to \bar f] &\propto &{1\over 2} e^{-\Gamma t}
|\bar A_{\bar f}|^2        \left\{
[1 + |\rho_{\bar f}|^2] +  [1 - |\rho_{\bar f}|^2] \cos{(\Delta M t)}
\right. \quad\\ \nonumber && \qquad\qquad\quad\left.
  - 2 \mbox{\rm Im}\left( {p\over q} \rho_{\bar f}\right) \sin{(\Delta M t)}
        \right\} ,
 \eeqn
where we have introduced the notation
\be
\begin{array}{lll}
A_f \equiv A[B^0\to f] , \quad & \bar A_f \equiv -A[\bar B^0\to f] , \quad &
\bar\rho_f\equiv \bar A_f / A_f , \quad
\\
A_{\bar f} \equiv A[B^0\to \bar f], \quad &
\bar A_{\bar f} \equiv -A[\bar B^0\to \bar f] , \quad &
\rho_{\bar f}\equiv A_{\bar f} / \bar A_{\bar f} . \quad
\ea
\ee

CP invariance demands the probabilities of CP conjugate processes to be
identical.
Thus, CP conservation requires
$A_f = \bar A_{\bar f}$, $A_{\bar f} = \bar A_f$,
$\bar\rho_f = \rho_{\bar f}$ and
${\rm Im}({q\over p} \bar\rho_f) = {\rm Im}({p\over q} \rho_{\bar f})$.
Violation of any of the first three equalities would be a signal of
direct CP violation. The fourth equality tests CP violation generated
by the interference of the direct decay $B^0\to f$ and the
mixing-induced decay $B^0\to\bar B^0\to f$.

Note that in order to be able to observe any CP-violating asymmetry,
one needs to distinguish between $B^0$ and $\bar B^0$ decays.
However, a final state $f$ that is common to both $B^0$ and $\bar B^0$ decays
cannot reveal by itself whether it came from a $B^0$ or a $\bar B^0$.
Therefore, one needs independent information
on the flavour identity of the decaying neutral $B$ meson;
this is referred to as ``flavour tagging''.
Since beauty hadrons are always produced in pairs, one can use for
instance the flavour-specific decays of one $B$ to ``tag'' the
flavour of the companion $B$.

An obvious example of final states $f$ which can be reached both from the
$B^0$ and the $\bar B^0$ are CP eigenstates, i.e. states such that
$\bar f = \zeta_f f$  ($\zeta_f = \pm 1$).
The ratios $\bar\rho_f$ and $\rho_{\bar f}$ depend in general on the
underlying strong dynamics.
However,
for CP self-conjugate final states, all dependence on the
strong interaction disappears \cite{CS,BS}
if only one weak amplitude contributes to
the $B^0\to f$ and $\bar B^0\to f$ transitions.
In this case, we can write the decay amplitude as
$A_f = M e^{i \phi_D} e^{i \delta_s}$, where $M = M^*$, $\phi_D$ is the phase
of the weak decay amplitude and $\delta_s$ is the strong phase associated with
final-state interactions.
It is easy to check that the ratios $\bar\rho_f$ and $\rho_{\bar f}$
are then given  by
($A_{\bar f} = M \zeta_f e^{i\phi_D} e^{i\delta_s}$,
 $\bar A_{f} = M  \zeta_f e^{-i\phi_D} e^{i\delta_s}$,
 $\bar A_{\bar f} = M  e^{-i\phi_D} e^{i\delta_s}$)
\be
\rho_{\bar f} = \bar\rho_f^* = \zeta_f e^{2i\phi_D} .
\ee
The unwanted effect of final-state interactions cancels out completely
from these two ratios.
Moreover, $\rho_{\bar f}$ and $\bar\rho_f$ simplify in this case to
a single weak phase, associated with the underlying weak quark transition.

Since $|\rho_{\bar f}| = |\bar\rho_f| = 1$,
the time-dependent decay probabilities given in Eqs. (\ref{eq:decay_b})
and (\ref{eq:decay_bbar}) become much simpler. In particular, there is no
longer any dependence on $\cos{(\Delta M t)}$.
Moreover, for $B$ mesons
$|\Gamma_{12}/M_{12}|<<1$, implying
\be
{q\over p} \,\approx\, \sqrt{{M_{12}^*\over M_{12}}} \,\approx\,
{V_{tb}^* V_{tq}^{\phantom{*}} \over V_{tb}^{\phantom{*}} V_{tq}^*}
\,\equiv\, e^{-2 i \phi_M} .
\ee
Here $q \equiv d, s$ stands for $B^0_d$, $B^0_s$.
In deriving this relation we have used the fact that $M_{12}$ is
dominated by the top contribution, due to the quadratic dependence with
the mass of the quark running along the internal lines of the box diagram.
Therefore, the mixing ratio $q/p$ is also given by a known weak phase,
and the coefficients of the sinusoidal terms in the time-dependent decay
amplitudes are then fully known in terms of CKM mixing angles only:
\be
\label{eq:im_coeff}
\mbox{\rm Im}\left( {p\over q} \rho_{\bar f}\right) \,\approx\,
-\mbox{\rm Im}\left( {q\over p} \bar\rho_f\right) \,\approx\,
\zeta_f\sin{[2(\phi_M + \phi_D)]}
\,\equiv\, \zeta_f\sin{(2\Phi)}.
\ee

The time-dependent decay rates are finally given by
\beqn
\label{eq:rate_b}
\Gamma[B^0(t)\to f] & = & \Gamma[B^0\to f] \, e^{-\Gamma t} \,
       \{ 1 + \zeta_f\sin{(2 \Phi)} \sin{(\Delta M t)} \} , \\
\label{eq:rate_bbar}
\Gamma[\bar B^0(t)\to \bar f] & = & \Gamma[\bar B^0\to \bar f]
   \, e^{-\Gamma t} \,
       \{ 1 - \zeta_f\sin{(2 \Phi)} \sin{(\Delta M t)} \} .
\eeqn

\noindent In this ideal case,
the time-dependent CP-violating decay asymmetry
\be
{\Gamma[B^0(t)\to f] - \Gamma[\bar B^0(t)\to\bar f] \over
 \Gamma[B^0(t)\to f] + \Gamma[\bar B^0(t)\to\bar f]} \, = \,
 \zeta_f\sin{(2 \Phi)} \, \sin{(\Delta M t)}
\ee
provides a direct and clean measurement of  the CKM parameters
\cite{KLPS}.
Integrating over all decay times yields
\be
\label{eq:integ_rate}
\int_0^\infty dt \,\Gamma[\BBB(t)\to\ffb] \,\propto\,
1 \mp\zeta_f\,\sin{(2 \Phi)}
  \, {x \over 1 + x^2} ,
\ee
where $x\equiv\Delta M/\Gamma$.
For $B^0_d$ mesons, $x_d = 0.70\pm 0.07$  \cite{mixing}; thus, the mixing
term $x_d/(1+x_d^2)$ suppresses the observable asymmetry by a factor of about
two.
For $B^0_s$ mesons, one expects
$x_s \sim x_d |V_{ts}|^2/|V_{td}|^2\sim x_d/\{ \lambda^2 [(1-\rho)^2+\eta^2]\}
 >> x_d$, and
therefore the large $B^0_s$-$\bar B^0_s$ mixing would
lead to a huge dilution of the CP asymmetry.
The measurement of the time-dependence is then a crucial requirement for
observing CP-violating asymmetries with $B^0_s$ mesons.

In $e^+e^-$ machines, running near the $B^0\bar B^0$ production threshold,
there is an additional complication coming from the fact that the $B$
meson used to ``tag'' the flavour is also a neutral one, and therefore
both mesons oscillate.
Moreover, the $B^0\bar B^0$ pair is produced in a coherent quantum
state which is a C eigenstate (C-odd in $e^+e^-\to B^0\bar B^0$,
C-even in  $e^+e^-\to B^0\bar B^0\gamma$).
Taking that into account, the observable time-dependent
asymmetry takes the form
\beqn
{\Gamma[(B^0\bar B^0)_{C=\mp}\to f + (l^-\bar\nu_l X^+)]
 - \Gamma[(B^0\bar B^0)_{C=\mp}\to f + (l^+\nu_l X^-)] \over
\Gamma[(B^0\bar B^0)_{C=\mp}\to f + (l^-\bar\nu_l X^+)]
+\Gamma[(B^0\bar B^0)_{C=\mp}\to f + (l^+\nu_l X^-)]}
 \, =\qquad\qquad\quad &&
\nonumber\\
\qquad\qquad\qquad\qquad\qquad\qquad\qquad\qquad\quad\quad
 \zeta_f \sin{(2\Phi)} \sin{[\Delta M (t\mp \bar t)]}\, ,\,\,
\eeqn
where the $B$ flavour has been assumed to be ``tagged'' through
the semileptonic decay, and
$t$ ($\bar t$) denotes the time of decay into $f$ ($l^\pm$).
Note that for $C=-1$ the asymmetry vanishes if $t$ and $\bar t$ are
treated symmetrically.
A measurement of at least the sign of $\Delta t \equiv t - \bar t$
is necessary to detect CP violation in this case.

\begin{table}[tbh]
\centering
\hphantom{}
\begin{tabular}{|c|c|c|l|c|}
\hline
Decay & CKM factor & CKM factor & \,\,\,\, Exclusive channels &
 $\quad \Phi \quad$ \\
           & (Direct) & (Penguin) & & \\ \hline
$\bar b \to \bar c  c \bar s$ & $A \lambda^2$ & $-A \lambda^2$ &
$ B^0_d\to J/\psi K_S ,
J/\psi K_L$ & $\beta$ \\
&&& $ B^0_s\to D_s^+ D_s^-, J/\psi\eta$ & 0 \\ \hline
$\bar b\to\bar s s \bar s$ & -- & $-A \lambda^2$ &
$ B^0_d\to K_S\phi, K_L\phi$ &
$\beta$ \\ \hline
$\bar b\to\bar d d\bar s$ & -- & $-A \lambda^2$ &
$ B^0_s\to K_S K_S, K_L K_L$ &
0 \\ \hline
$\bar b\to\bar c c\bar d$ & $-A\lambda^3$ & $A\lambda^3 (1-\rho - i \eta)$ &
$ B^0_d\to D^+ D^- , J/\psi\pi^0$ & $\approx \beta$
\\ &&& $ B^0_s\to J/\psi K_S, J/\psi K_L$ & 0 \\ \hline
$\bar b\to\bar u u\bar d$ & $A \lambda^3 (\rho + i \eta)$ &
$A\lambda^3 (1 - \rho - i \eta)$ &
$ B^0_d\to\pi^+\pi^- , \rho^0\pi^0 ,\omega\pi^0$
& $\approx \beta+\gamma$ \\
 & & & $ B^0_s\to\rho^0 K_S ,\omega K_S ,\pi^0 K_S$,
  & $\approx \gamma$ \\
 & & & $\phantom{ B^0_s\to} \rho^0 K_L, \omega K_L, \pi^0 K_L$ &
\\ \hline
$\bar b\to\bar s s\bar d$ & -- & $A\lambda^3 (1 - \rho - i \eta)$ &
$ B^0_d\to K_S K_S, K_L K_L$ & 0
\\ &&& $ B^0_s\to K_S\phi, K_L\phi$ & $-\beta$
\\ \hline
\end{tabular}
\caption{CKM factors and relevant angle $\Phi$ for some $B$-decays into
CP-eigenstates.}
\label{tab:decays}
\end{table}

We have assumed up to now that there is only one amplitude contributing
to the given decay process.
Unfortunately, this is usually not the case.
If several decay amplitudes
with different weak and strong phases
contribute, $|\bar{\rho}_f|\not=1$, and the interference term will
depend both on the CKM mixing parameters and on the strong dynamics embodied
in the ratio $\bar{\rho}_f$.

The leading contributions to
$\bar b\to\bar q' q'\bar q$ decay amplitudes are either
``direct'' (Fermi) or  generated by gluon exchange (``penguin'').
Although of higher order in the strong coupling constant,  penguin
amplitudes are logarithmically enhanced, due to the virtual $W$-loop, and
are  potentially competitive. Table \ref{tab:decays} contains the CKM
factors associated with the direct and penguin diagrams for
different $B$-decay modes into CP-eigenstates.
Also shown is the relevant angle $\Phi$.
In terms of CKM elements, the angles $\alpha$, $\beta$
and $\gamma$ are:
\be\label{eq:angles}
\alpha\equiv\arg{\left[
   -{V^{\phantom{*}}_{td}V^*_{tb}\over V^{\phantom{*}}_{ud}V^*_{ub}}
  \right]} , \quad
\beta\equiv\arg{\left[
   -{V^{\phantom{*}}_{cd}V^*_{cb}\over V^{\phantom{*}}_{td}V^*_{tb}}
  \right]} , \quad
\gamma\equiv\arg{\left[
   -{V^{\phantom{*}}_{ud}V^*_{ub}\over V^{\phantom{*}}_{cd}V^*_{cb}}
  \right]} , \quad
\ee
Due to the unitarity of the CKM matrix,
$\alpha + \beta + \gamma = \pi$.

The $\bar b\to\bar c c\bar s$ quark decays are theoretically unambiguous
\cite{LP}:
the direct and penguin amplitudes have the same
weak phase $\Phi = \beta $ ($0$), for $ B^0_d $ ($ B^0_s$). Ditto for
$\bar b\to\bar s s\bar s$ and $\bar b\to\bar d d\bar s$,
where only the penguin mechanism is possible.
The same is true for the Cabibbo-suppressed $\bar b\to\bar s s \bar d$ mode,
which only gets contribution from the penguin diagram;
the $ B^0_d$ ($ B^0_s$) phases are 0 ($-\beta$) in this case.
The $\bar b\to\bar c c\bar d$ and $\bar b\to\bar u u\bar d$ decay modes
are not so
simple; the two decay mechanisms
have the same Cabibbo suppression ($\lambda^3$) and
different weak phases,
but the
penguin amplitudes are  down by
${(\alpha_s / 6 \pi}) \ln(m_W / m_b) \approx 3 \% $:
these decay modes can be used
as approximate measurements of the CKM factors.
We have  not considered doubly Cabibbo-suppressed
decay amplitudes, such as
$\bar b\to\bar u u\bar s$, for which
penguin effects can be important and
spoil the simple estimates based on the direct decay mechanism.

Presumably  the most realistic channels for the measurement of the angles
  $\Phi=(\beta ,\,\alpha ,\,\gamma)$ are
   $ B^0_d\to J/\psi K_S$, $ B^0_d\to\pi^+\pi^-$
($\beta + \gamma = \pi - \alpha$)
and $ B^0_s\to\rho^0 K_S$,
respectively. The first of these processes is no doubt
the one with the cleanest signature and the most tractable background.
The last process has the disadvantage of requiring a $B^0_s$ meson and,
moreover,
its branching ratio is expected to be very small because the
``direct'' decay amplitude is colour suppressed, leading presumably to
a much larger penguin contamination;
thus, the determination
of $\gamma$, through this decay mode looks a quite formidable task.

 The decay modes where $\Phi = 0$ are useless for making
a determination of the CKM factors.
However, they provide a very interesting test
of the SM mechanism of CP-violation, because the prediction
that no CP-asymmetry should be seen for these modes is very clean.
Any detected CP-violating signal would be a clear indication of new physics.

Many other decay modes of $B$ mesons can be used to get information on
the CKM factors responsible for CP violation phenomena. A recent
summary, including alternative ways of measuring $\gamma$, can be found
in ref. \cite{ref:ecfa}.

\section{Summary}
\label{sec:summary}

The SM incorporates a mechanism to generate CP violation, through the
single phase naturally occurring in the CKM matrix.
So far, only one non-zero CP-violation effect has been clearly
established: $\varepsilon \approx 2.3\times 10^{-3} e^{i\pi/4}$.
Therefore, we do not have yet an experimental test of the CKM mechanism.
The value of $\varepsilon$ is just fitted with the CKM phase; but we
could also fit this measured parameter using other non-standard
sources of CP violation.
In fact, the present observations can still be explained with the
old ``superweak'' mechanism \cite{wo:1}, which associates
CP violation with some unknown $\Delta S=2$ interaction, i.e. with
a $K^0$-$\bar K^0$ mixing effect.

Since all CP-violating effects are supossed to be generated by the
same CKM phase, the SM predictions are quite constrained. Moreover,
as shown in Sect. \ref{sec:SM}, the CKM mechanism implies very
specific requirements for CP-violation phenomena to show up.
The experimental verification of the SM predictions is obviously a
very important challenge for future experiments, which could lead
to big surprises.

In the SM, CP violation is associated with a charged-current interaction
which changes the quark flavours in a very definite way:
$u_i\to d_j W^+$, $d_j\to u_i W^-$. Therefore, CP should be
directly violated in many ($\Delta S=1$, $\Delta D=1$, $\Delta B=1$, \ldots)
decay processes without any relation with meson-antimeson mixing.
Although the quantitative predictions are often uncertain,
owing to the not so-well understood long-distance strong-interaction
dynamics, the experimental observation of a non-zero
CP-violating asymmetry in any self-tagging decay mode would be a major
achivement, as it would clearly establish the existence of
direct CP violation in the decay amplitudes.

The SM mechanism of CP violation is based in the unitarity of the
CKM matrix. Testing the unitarity relations of the CKM matrix elements
is then a way to test the source of CP violation.
Up to now, the only relation which has been precisely tested
is the one
associated with the
first row of the CKM matrix:
\be\label{eq:utest}
|V_{ud}|^2 + |V_{us}|^2 + |V_{ub}|^2 \, = \, 0.9981\pm0.0027 .
\ee
The unitarity relation is very well satisfied in this case,
providing a nice confirmation of the SM\footnote{
%
In fact, this is a test of the SM radiative corrections, which are
crucial for extracting $V_{ud}$ with the quoted precision.
If these corrections were neglected, unitarity would be violated by
many $\sigma$'s \protect\cite{ref:marciano}.}.
%
However, only the moduli of the CKM matrix elements are involved in
Eq. (\ref{eq:utest}), while CP violation has to do with their phases.

More interesting are the off-diagonal unitarity conditions:
\be\label{eq:triangles}
\matrix{\displaystyle V^\ast_{ ud}V^{\phantom{*}}_{us}
& \displaystyle + & \displaystyle
V^\ast_{ cd}V^{\phantom{*}}_{cs}
& \displaystyle + & \displaystyle V^\ast_{ td}V^{\phantom{*}}_{ts} &
\displaystyle = & \displaystyle 0 \, ,
\cr\displaystyle  & \displaystyle  &
\displaystyle  & \displaystyle  & \displaystyle  & \displaystyle  &
\displaystyle \cr\displaystyle V^\ast_{ us}V^{\phantom{*}}_{ub}
& \displaystyle + &
\displaystyle V^\ast_{ cs}V^{\phantom{*}}_{cb}
& \displaystyle + & \displaystyle
V^\ast_{ ts}V^{\phantom{*}}_{tb} & \displaystyle = & \displaystyle 0 \, ,
\cr\displaystyle  &
\displaystyle  & \displaystyle  & \displaystyle  & \displaystyle  &
\displaystyle  & \displaystyle \cr
\displaystyle V^\ast_{ ub}V^{\phantom{*}}_{ud} &
\displaystyle + & \displaystyle V^{\ast  }_{cb}V^{\phantom{*}}_{cd}
& \displaystyle + &
\displaystyle V^\ast_{ tb}V^{\phantom{*}}_{td}
& \displaystyle = & \displaystyle 0\, .
 \cr}
\ee
These relations can be visualized by triangles in a complex
plane \cite{bj:1}.
Owing to Eq.~(\ref{eq:J_relation}), the three triangles have the
same area $|\cJ|/2$.
In the absence of CP violation, these triangles would degenerate
into segments along the real axis.

In the first two triangles, one side is much shorter than the other
two (the Cabibbo suppression factors of the three sides are
$\lambda$, $\lambda$ and $\lambda^5$ in the first triangle,
and $\lambda^4$, $\lambda^2$ and $\lambda^2$ in the second one).
This is the reason why CP effects are so small for $K$ mesons
(first triangle), and why certain  asymmetries in $B_s$ decays are
predicted to be tiny (second triangle).

\begin{figure}[htb]
\vfill
\caption{The unitarity triangle. Also shown are various topics in  $B$
physics that allow to measure its sides and angles
\protect\cite{ref:ecfa}.}
\label{fig:utriangle}
\vfill
\end{figure}

The third triangle looks more interesting, since the
three sides have a similar size of about $\lambda^3$.
They are small, which means that the relevant $b$-decay branching ratios
are small, but once enough $B$ mesons would be produced, CP-violation
asymmetries are going to be sizeable.
This triangle is shown in Fig. \ref{fig:utriangle}, where it has
been scaled by dividing its sides by
$|V^{\ast  }_{cb}V^{\phantom{*}}_{cd}|$.
In the Wolfenstein parametrization (\ref{eq:wolfenstein}),
where $V^{\ast  }_{cb}V^{\phantom{*}}_{cd}$ is real, this aligns one
side of the triangle along the real axis and makes its length equal to
1; the coordinates of the 3 vertices are then
$(0,0)$, $(1,0)$ and $(\rho,\eta)$.
The three angles of the triangle
are just the angles $\alpha$, $\beta$ and $\gamma$ in
Eq. (\ref{eq:angles}), which regulate the $B$-decay asymmetries.
Note that, although the orientation of the triangle in the complex plane
is phase-convention dependent, the triangle itself is a physical
object: the length of the sides and/or the angles can be directly
measured.

At present, we already have some information on this unitarity triangle.
CP-conserving measurements can provide a determination of its sides.
The measured \cite{mixing}
$\Gamma(b\to u)/\Gamma(b\to c)$ ratio fixes one side to be:
\be
R_b\equiv \Big| {V_{ub}\over\lambda V_{cb}}\Big| = \sqrt{\rho^2 +\eta^2}
= 0.34\pm 0.12 .
\ee
The other side can be extracted from the observed \cite{mixing}
$B^0_d$-$\bar B^0_d$ mixing:
\be\label{eq:rt}
R_t\equiv\Big| {V_{td}\over\lambda V_{cb}}\Big| =
\sqrt{(1-\rho)^2 +\eta^2} =
0.97\pm 0.43  .
\ee
$R_t$ depends quite sensitively on the non-perturbative
parameter $B_B f_B^2$
[the $B$ analogous of $B_K f_K^2$ in Eq. (\ref{eq:bk})] and on $m_t$.
The number in Eq. (\ref{eq:rt}) corresponds to the presently favoured
values $\sqrt{B_B} f_B = (1.6\pm0.4) f_\pi$ and $m_t = 160\pm 30$.
In principle, the measurement of these two sides
could make possible to establish that CP is violated
(assuming unitarity), by showing that they indeed give rise to a
triangle and not to a straight line.
With the present experimental and theoretical errors, this is however
not possible.

A third constraint is obtained from the measured value of
the CP-violation parameter $\varepsilon$, which forces the $(\rho,\eta)$
vertex to lie in the hyperbola (\ref{eq:hyperbola}).
The value of the $B_K$ parameter is the dominant source of uncertainty;
taking the conservative estimate $1/3 < B_K <1$, which is in agreement
with all determinations in Table~\ref{tab:bk}, $\eta>0$, but the sign of
$\rho$ is still not fixed.
The final allowed domain for the vertex $(\rho,\eta)$, satisfying these
three constraints,
is quite large because of the present theoretical uncertainties.

The observation of
CP-violating asymmetries with neutral $B$ mesons,
would allow to independently measure the three angles of
the triangle, providing an overconstrained determination
of the CKM matrix.
As shown in the previous section,
theoretical uncertainties can largely be avoided, for instance in decays
into CP-selfconjugate states,
so that CP-odd signals can be directly translated into
clean measurements of these angles.
If the measured sides and angles turn out to be consistent
with a geometrical triangle, we would have a beautiful test
of the CKM unitarity, providing strong support to the SM
mechanism of CP violation.
On the contrary, any deviation from a triangle shape
would be a clear proof that new physics is needed in order
to understand  CP-violating phenomena.

CP violation is a broad and fascinating subject, which is closely
related to the so-far untested scalar sector of the SM.
New experimental facilities are needed
for exploring CP-violating phenomena and test the SM predictions.
Large surprises may well be discovered, probably giving the first
hints of new physics and offering clues to the problems of
fermion-mass generation, quark mixing and family replication.

\begin{Thebibliography}{99}

\bibitem{ko:1}
M. Kobayashi and T. Maskawa
\Journal Prog. Theor. Phys.&42&652(1973).

\bibitem{ca:1} N. Cabibbo \Journal\PRL&10&531(1963).

\bibitem{ref:CP_book}
   {\em CP Violation}, ed. C. Jarlskog,
  Advanced Series on Directions in High Energy Physics
(World Scientific, Singapore, 1989), Vol. 3.

\bibitem{ref:CP_reviews}
   B. Winstein and L. Wolfenstein, {\em The Search for Direct CP Violation},
     EFI 92-55 (to appear in Rev. Mod. Phys.);\\
   W. Grimus, Fortsch. der Physik 36 (1988) 201;\\
   J.F. Donoghue, B.R. Holstein and G. Valencia, Int. J. Mod. Phys. A2
      (1987) 319.

\bibitem{ref:bh92} A.J. Buras ans M.K. Harlander, {\em A Top Quark Story:
  Quark Mixing, CP Violation and Rare Decays in the Standard Model}, in
{\em Heavy Flavours}, ed. A. J. Buras and M. Lindner,
Advanced Series on Directions in High Energy Physics
(World Scientific, Singapore, 1992), Vol. 10, p.~58.

\bibitem{ref:nakada}
  T. Nakada, {\em Review on CP Violation}, talk at the XVI International
      Symposium on Lepton-Photon Interactions at High Energies (Cornell,
    1993), preprint PSI-PR-93-18.

\bibitem{ref:CP_talks}
   R. Aleksan, {\em CP Violation}, Talk at the XXI International Meeting
      on Fundamental Physics (Miraflores de la Sierra, Spain, 1993),
      Saclay preprint DAPNIA/SPP 93-10;\\
   Y. Nir, {\em CP Violation}, Lectures at the $20^{\mbox{\rms th}}$
      Annual Summer Institute on Particle Physics (SLAC, 1992),
      preprint SLAC-PUB-5874.

\bibitem{ref:PDG92} Particle Data Group,
{\em Review of Particle Properties}, Phys. Rev. D45 (1992) Part~2.

\bibitem{wo:2} L. Wolfenstein \Journal\PRL&51&1945(1983).

\bibitem{ref:jarlskog} C. Jarlskog, Phys. Rev. Lett. 55 (1985) 1039;
     Z. Phys. C29 (1985) 491.

\bibitem{ref:NA31_ep} G.D. Barr et al., Phys. Lett. B317 (1993) 233;\\
       H. Burkhardt et al., Phys. Lett. B206 (1988) 169.

\bibitem{ref:E731_ep} L.K. Gibbons et al., Phys. Rev. Lett. 70 (1993) 1203.

\bibitem{ref:cern} G. Baar et al., CERN/SPSC/90-22 (1990).

\bibitem{ref:fnal} K. Arisaka et al., Proposal E832 (1990).

\bibitem{ref:DAPHNE} {\em The DA$\Phi$NE Physics Handbook},
   eds. L. Maiani, G. Pancheri and N. Paver (Frascati, 1992).

\bibitem{ref:GIM} S.L. Glashow, J. Iliopoulos and L. Maiani,
    Phys. Rev. D2 (1970) 1285.

\bibitem{ref:dgh82} J.F. Donoghue, E. Golowich and B.R. Holstein,
  Phys. Lett. 119B (1982) 412.

\bibitem{ref:pr85}
  A. Pich and E. de Rafael, Phys. Lett. 158B (1985) 477;\\
  J. Prades, C.A. Dom\'{\i}nguez, J.A. Pe\~narrocha,
 A. Pich and E. de Rafael, Z. Phys. C51 (1991) 287.

\bibitem{ref:bdg88} N. Bilic, C.A. Dom\'{\i}nguez and B. Guberina,
   Z. Phys. C39 (1988) 351; and references therein.

\bibitem{ref:pr91a} A. Pich and E. de Rafael, Nucl. Phys. B358 (1991) 311.

\bibitem{ref:bruno} C. Bruno, {\em Chiral Corrections to the
  $K^0$-$\bar K^0$ $B_K$-Parameter}, Oxford preprint OUTP-93-25P.

\bibitem{ref:bbg88}
 W.A. Bardeen, A.J. Buras and J.-M. G\'erard, Phys. Lett. 211B (1988) 343;\\
 J.-M. G\'erard, Proc. XXIV Int. Conf. on High Energy Physics, Munich, 1988,
eds. R. Kotthaus and J.H. K\"uhn, p.840.

\bibitem{ref:lattice}  C.T. Sachrajda, {\em Lattice Phenomenology},
    plenary talk at the EPS-93 Conference (Marseille, 1993);
   and references therein.

\bibitem{ref:ep}
  G. Buchalla, A.J. Buras and M.K. Harlander, Nucl. Phys.
     B337 (1990) 313; \\
  A.J. Buras, M. Jamin and M.E. Lautenbacher, Nucl. Phys. B408 (1993) 209;\\
  M. Ciuchini, E. Franco, G. Martinelli and L. Reina, Phys. Lett. B301
     (1993) 263.

\bibitem{ref:pr91b} A. Pich and E. de Rafael, Nucl. Phys. B367 (1991) 313;
  and references therein.

\bibitem{ref:edm}
  K.F. Smith et al., Phys. Lett. B234 (1990) 191;\\
  I.S. Altarev et al., JETP Lett. 44 (1987) 460.

\bibitem{ref:pq}
    R.D. Peccei and H.R. Quinn, Phys. Rev. Lett. 38 (1977) 1440;
       Phys. Rev. D16 (1977) 1791.

\bibitem{ref:k_rev} J.L. Ritchie and S.G. Wojcicki, {\em Rare K Decays},
Texas University preprint CPP-93-16 (to appear in Rev. Mod. Phys.);\\
   L. Littenberg and G. Valencia, {\em Rare and Radiative Kaon Decays},
    Annu. Rev. Nucl. Part. Sci. 43 (1993).

\bibitem{littenberg}
   L.S. Littenberg, Phys. Rev. D39 (1989) 3322.

\bibitem{ref:ddgb}
  G. Buchalla, A.J. Buras and M.K. Harlander, Nucl. Phys. B349 (1991) 1;
  \\
   C. Dib, I. Dunietz and F.J. Gilman, Phys. Lett. B218 (1989) 487;
     Phys. Rev. D39 (1989) 2639;
  \\ J. Flynn and L. Randall, Nucl. Phys. B326 (1989) 31.

\bibitem{ref:epr88}
   G. Ecker, A. Pich and E. de Rafael, Nucl. Phys. B303 (1988) 665;
  Nucl. Phys. B291 (1987) 692.

\bibitem{ref:mexico} A. Pich, {\em Introduction to Chiral Perturbation
   Theory}, CERN-TH.6978/93.

\bibitem{ref:epr87}
   G. Ecker, A. Pich and E. de Rafael, Phys. Lett. B189 (1987) 363;\\
   L. Cappiello and G. D'Ambrosio, Nuovo Cim. 99A (1988) 153.

\bibitem{ref:epr90} G. Ecker, A. Pich and E. de Rafael,
  Phys. Lett. B237 (1990) 481.

\bibitem{ref:NA31_92}
     G.D. Barr et al., Phys. Lett. B284 (1992) 440;
                 B242 (1990) 523.

\bibitem{ref:CEP93}
     A. Cohen, G. Ecker and A. Pich, Phys. Lett. B304 (1993) 347.

\bibitem{ref:BL86}
  F.J. Botella and C.S. Lim, Phys. Rev. Lett. 56 (1986) 1651.

\bibitem{ref:HIGGS}
    D. Treille, {\em Particle Searches},
     plenary talk at the EPS-93 Conference (Marseille, 1993).

\bibitem{ref:GN89}
      C.Q. Geng and J.N. Ng, Phys. Rev. D39 (1989) 3330.

\bibitem{ref:MO93}
      R.N. Mohapatra, Prog. Part. Nucl. Phys. 31 (1993) 39; \\
      C.Q. Geng and J.N. Ng, Phys. Rev. D42 (1990) 1509.

\bibitem{ref:EP91}
     G. Ecker and A. Pich, Nucl. Phys. B366 (1991) 189.

\bibitem{ref:HE83}
      P. Herczeg, Phys. Rev. D27 (1983) 1512.

\bibitem{ref:NA31_87}
     H. Burkhardt et al., Phys. Lett. B199 (1987) 139.

\bibitem{ref:ecfa} R. Aleksan et al., {\em CP Violation in the B Meson
  System and Prospects at an Asymmetric B Meson Factory}, in
  {\em Proc. of the ECFA Workshop on a European B-Meson Factory},
  ed. R. Aleksan and A. Ali, ECFA 93-151; and references therein.

\bibitem{il} T. Inami and C.S. Lim
   \Journal Prog. Theor. Phys.&65&297(1981).

\bibitem{BSS} M. Bander, D. Silverman and A. Soni
   \Journal Phys. Rev.&43&242(1979).

\bibitem{GH} J.M. G\'{e}rard and W.S. Hou, Phys. Lett. B253 (1991) 478;
    Phys. Rev. D43 (1991) 2909.

\bibitem{CS} A. Carter and A.I. Sanda \Journal \PRL&45&952(1980);
    Phys. Rev. D23 (1981) 1567.

\bibitem{BS} I.I. Bigi and A.I. Sanda, Nucl. Phys. B193 (1981) 85.

\bibitem{BJ81} J. Bernab\'eu and C. Jarlskog, Z. Phys. C8 (1981) 233.

\bibitem{Baryon}
    J.D. Bjorken \Journal \NPB\ (Proc. Suppl.)&11&325(1989); \\
    I.I. Bigi and B. Stech, in {\em Proceedings of the Workshop on High
    Sensitivity Beauty physics} at Fermilab, Nov. 1987, ed.
    A.J. Slaughter, N. Lockyer and M. Schmidt, p. 239;\\
    I. Dunietz, Z. Phys. C56 (1992) 129.

\bibitem{KLPS} P. Krawczyk et al., Nucl. Phys. B307 (1988) 19.

\bibitem{mixing}
   D. Besson, {\em B Weak Decays from Threshold Experiments},
     talk at the XVI International
      Symposium on Lepton-Photon Interactions at High Energies (Cornell,
    1993);\\
   M. Danilov, {\em B Physics},
   plenary talk at the EPS-93 Conference (Marseille, 1993).

\bibitem{LP} D. London and R.D. Peccei, Phys. Lett. B223 (1989) 257.

\bibitem{wo:1} L. Wolfenstein \Journal\PRL&13&562(1964).

\bibitem{ref:marciano} W. Marciano, Annu. Rev. Nucl. Part. Sci. 41
   (1991) 469.

\bibitem{bj:1} L.-L. Chau and W.-Y. Keung \Journal\PRL&53&1802(1984);\\
J.D. Bjorken, Fermilab Preprint, 1988 (unpublished);\\
C. Jarlskog and R. Stora, Phys. Lett. B208 (1988) 268;\\
J.L. Rosner, A.I. Sanda and M.P. Schmidt, in proceedings of the Workshop
on High Sensitivity Beauty physics at Fermilab, Fermilab, Nov. 11-14, 1987.

\end{Thebibliography}

\end{document}